\documentclass{appolb}
\usepackage{graphicx}

\begin{document}
\title{Phenomenology of the $ppK^+K^-$ system near threshold
\thanks{Presented at the II International Symposium on Mesic Nuclei, Cracow, September 22 - 25, 2013}%
}
\author{M. Silarski
\address{Institute of Physics, Jagiellonian University, PL-30-059 Cracow, Poland}
}
\maketitle
\begin{abstract}
In this article studies of the near threshold $pp\to ppK^+K^-$ reaction conducted with the COSY-11
and the ANKE detectors are reviewed. In particular recent investigations on the $K^+K^-$
final state interaction are revisited taking into account updated cross sections of the COSY-11
experiment. These studies resulted in the new value of $K^+K^-$ effective range
amounting to:
$\mathrm{Re}(b_{K^{+}K^{-}}) = -0.2^{+0.8_{stat}~+0.4_{sys}}_{-0.6_{stat}~-0.4_{sys}}~\mathrm{fm}$ and
$\mathrm{Im}(b_{K^{+}K^{-}}) = 1.2^{~+0.5_{stat}~+0.3_{sys}}_{~-0.3_{stat}~-0.3_{sys}}~\mathrm{fm}$.
The determined real and imaginary parts of the $K^+K^-$ scattering length were estimated to be:
$\left|\mathrm{Re}(a_{K^{+}K^{-}})\right| = 10^{~+17_{stat}}_{~-10_{stat}}~\mathrm{fm}$
and $\mathrm{Im}(a_{K^{+}K^{-}}) = 0^{~+37_{stat}}_{~-10_{stat}}~\mathrm{fm}$.

\end{abstract}
\section{Introduction}
\label{intro}
\begin{figure}
\centering
\includegraphics[width=0.4\textwidth]{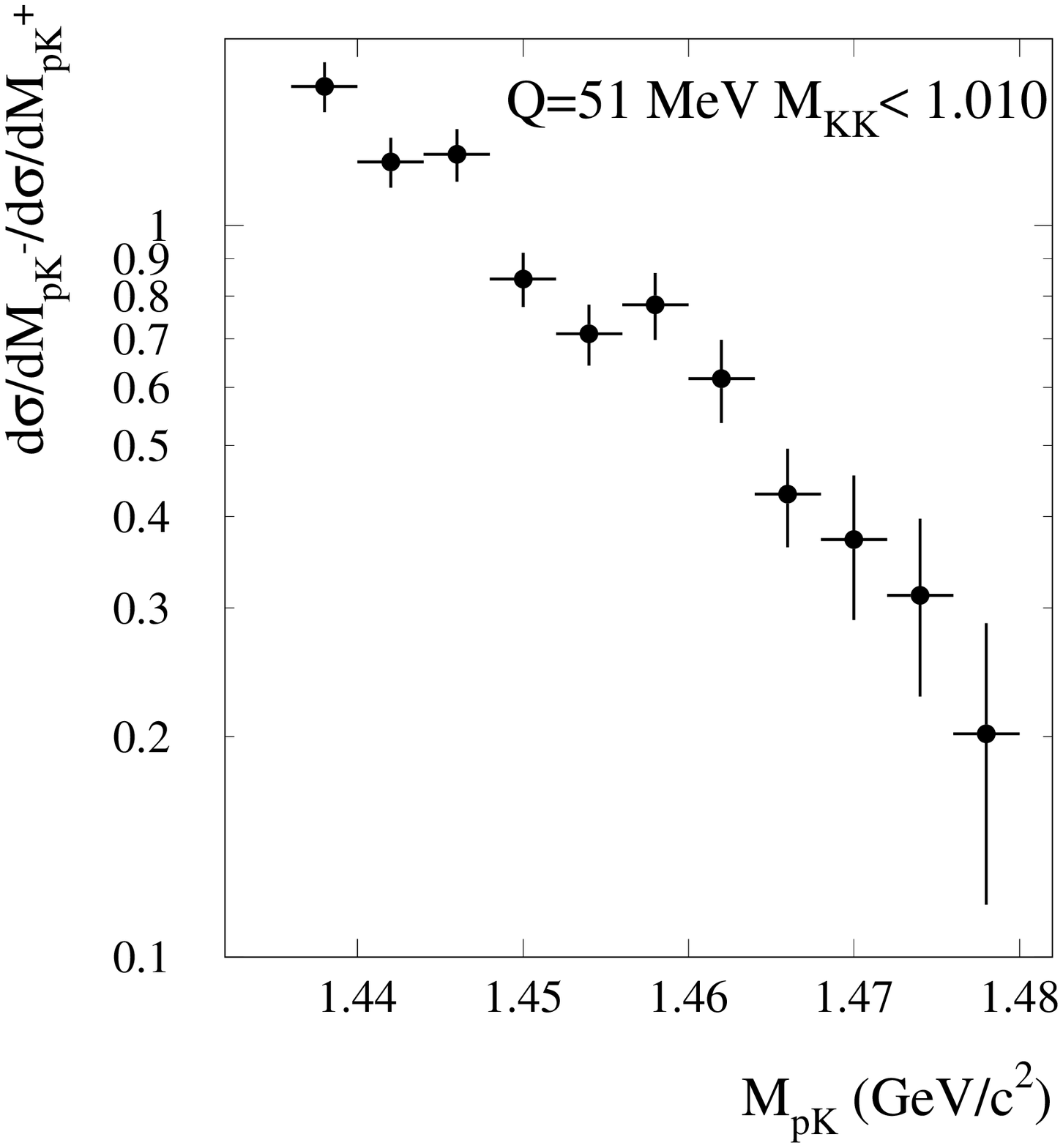}
\includegraphics[width=0.4\textwidth]{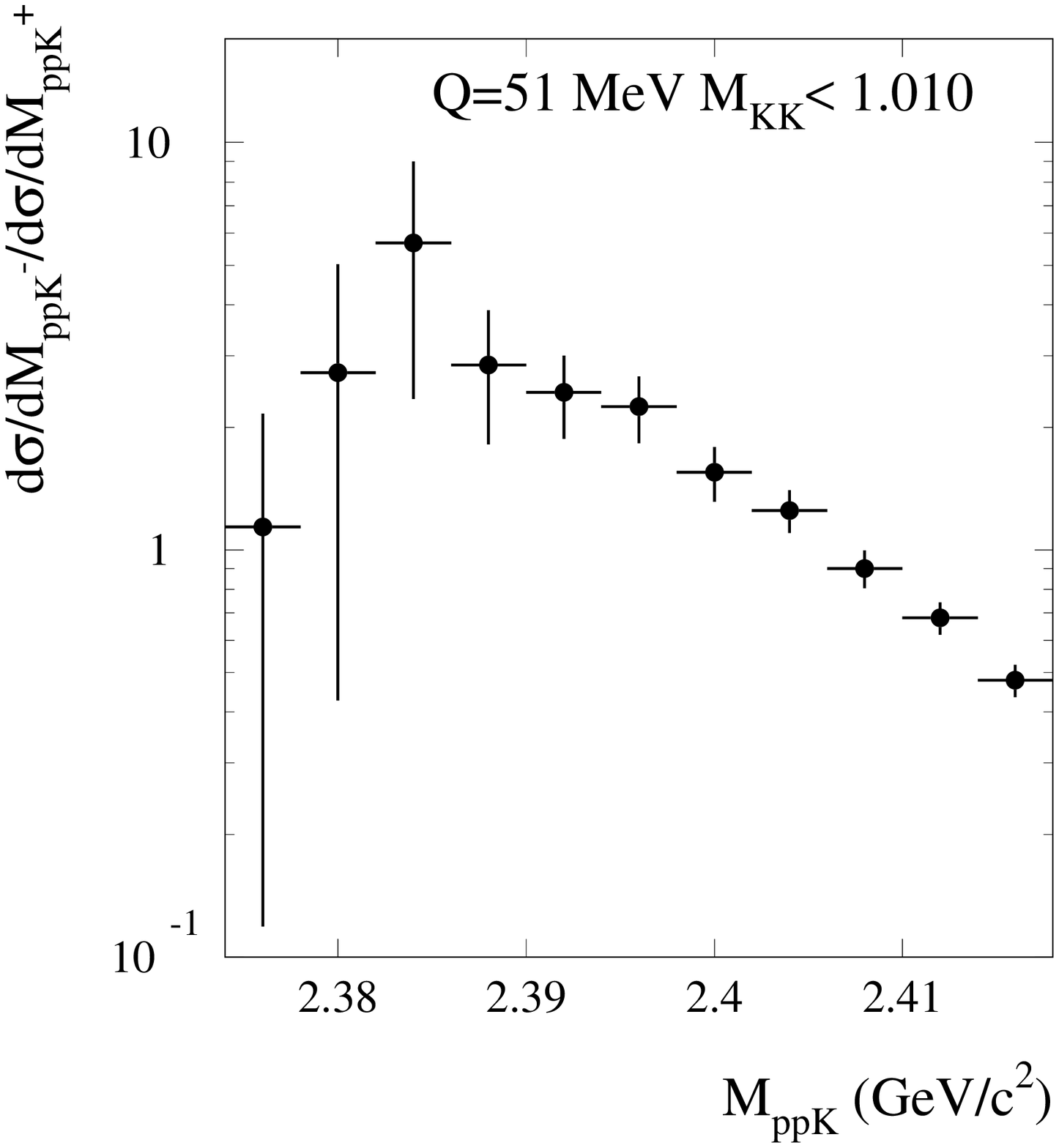}
\includegraphics[width=0.4\textwidth]{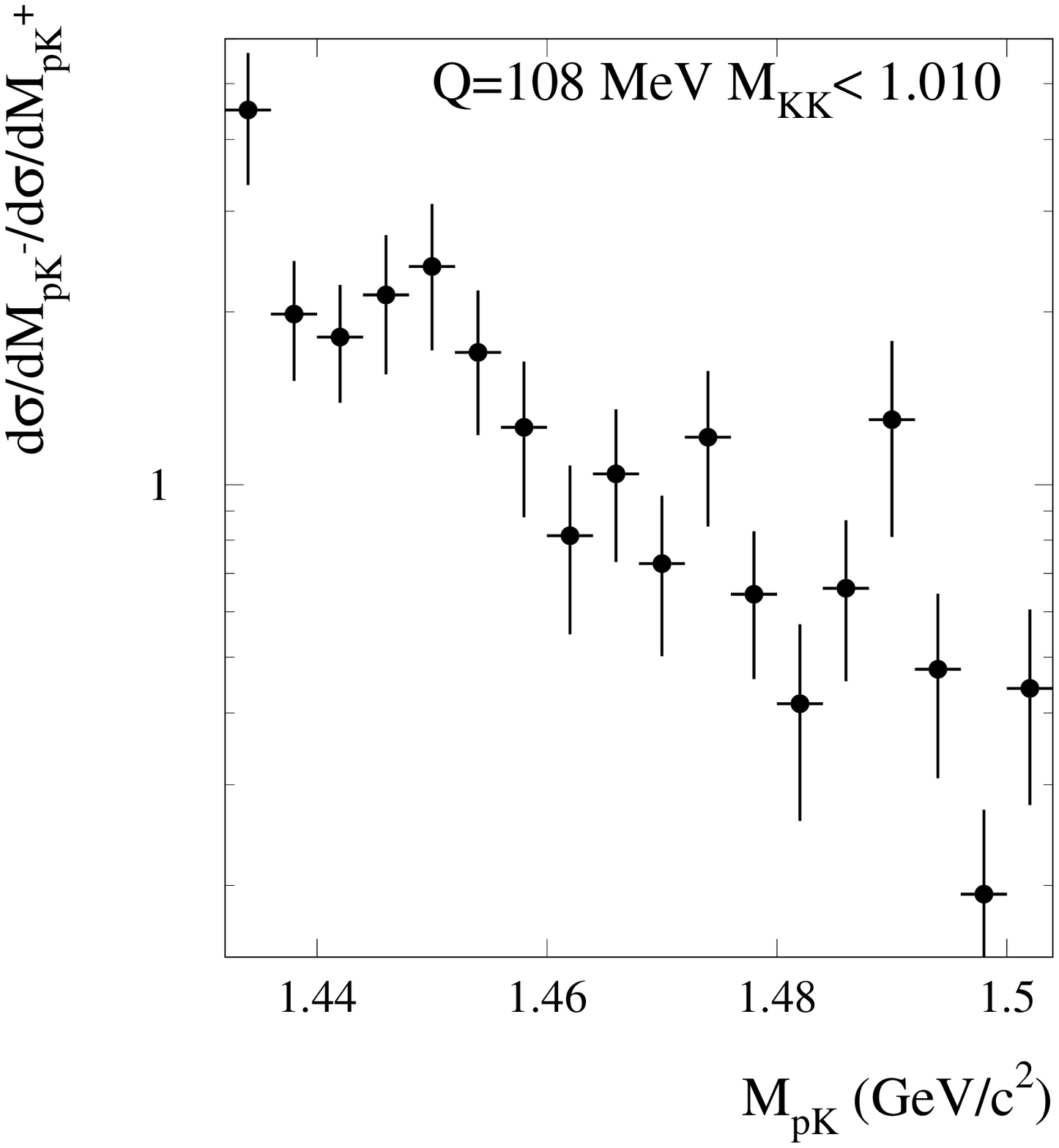}
\includegraphics[width=0.4\textwidth]{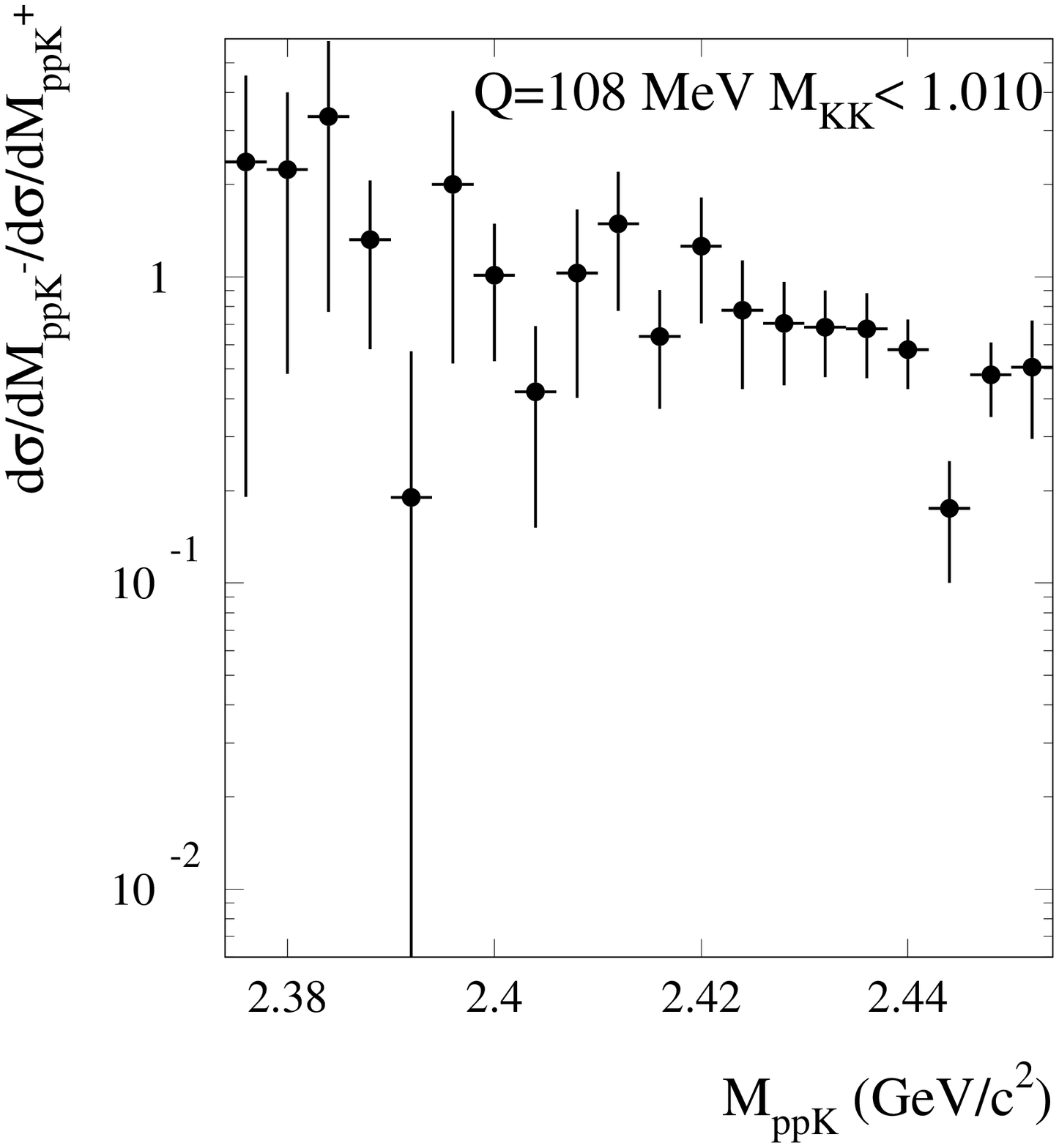}
\caption{Ratios of differential cross sections as a function of $pK^-$ invariant mass $M_{pK^{-}}$
and $ppK^{-}$ invariant mass $M_{ppK^{-}}$ measured by the ANKE experiment at excess energies
of Q = 51 MeV and Q = 108 MeV~\cite{anke}.}
\label{ratioA}
\end{figure}

The low energy $ppK^+K^-$ system provides opportunity to study both the $pK^-$ and $K^+K^-$
final state interactions. The latter is of a great importance in the still ongoing discussion
about the possible formation of the $K\bar{K}$ bound states~\cite{Lohse, Weinstein} which
requires a strong attractive potential. The $pK^-$ final state interaction (FSI) is also very
important in view of the unknown structure of the $\Lambda(1405)$ hyperon which is often considered
as the $NK^-$ molecule, and could provide some hints of existence of the deeply bound
$ppK^-$ kaonic states~\cite{Piscicchia,catalina}.\\
The dynamics of the $ppK^+K^-$ system has been studied mainly in the proton-proton collisions
at the cooler synchrotron COSY at the research center in J{\"u}lich, Germany~\cite{cosy}. COSY,
providing proton and deuteron beams with low emittance and small momentum spread, is an ideal
facility for measurements at threshold where the cross sections rise rapidly.
First measurements of the $pp\to ppK^+K^-$ reaction were performed by the COSY-11 collaboration
to study the properties of $f_{0}$ and $a_{0}$ scalar resonances which are proposed to be a bound
state of $K^+$ and $K^-$ mesons\footnote{Besides that interpretation these particles
were also considered to be ordinary $q\bar{q}$ mesons~\cite{Morgan}, tetraquark states~\cite{Jaffe},
hybrid $q\bar{q}$/meson-meson systems~\cite{Beveren} or even gluballs~\cite{Johnson}.}
\cite{Lohse, Weinstein}.
\begin{figure}
\centering
\includegraphics[width=0.49\textwidth]{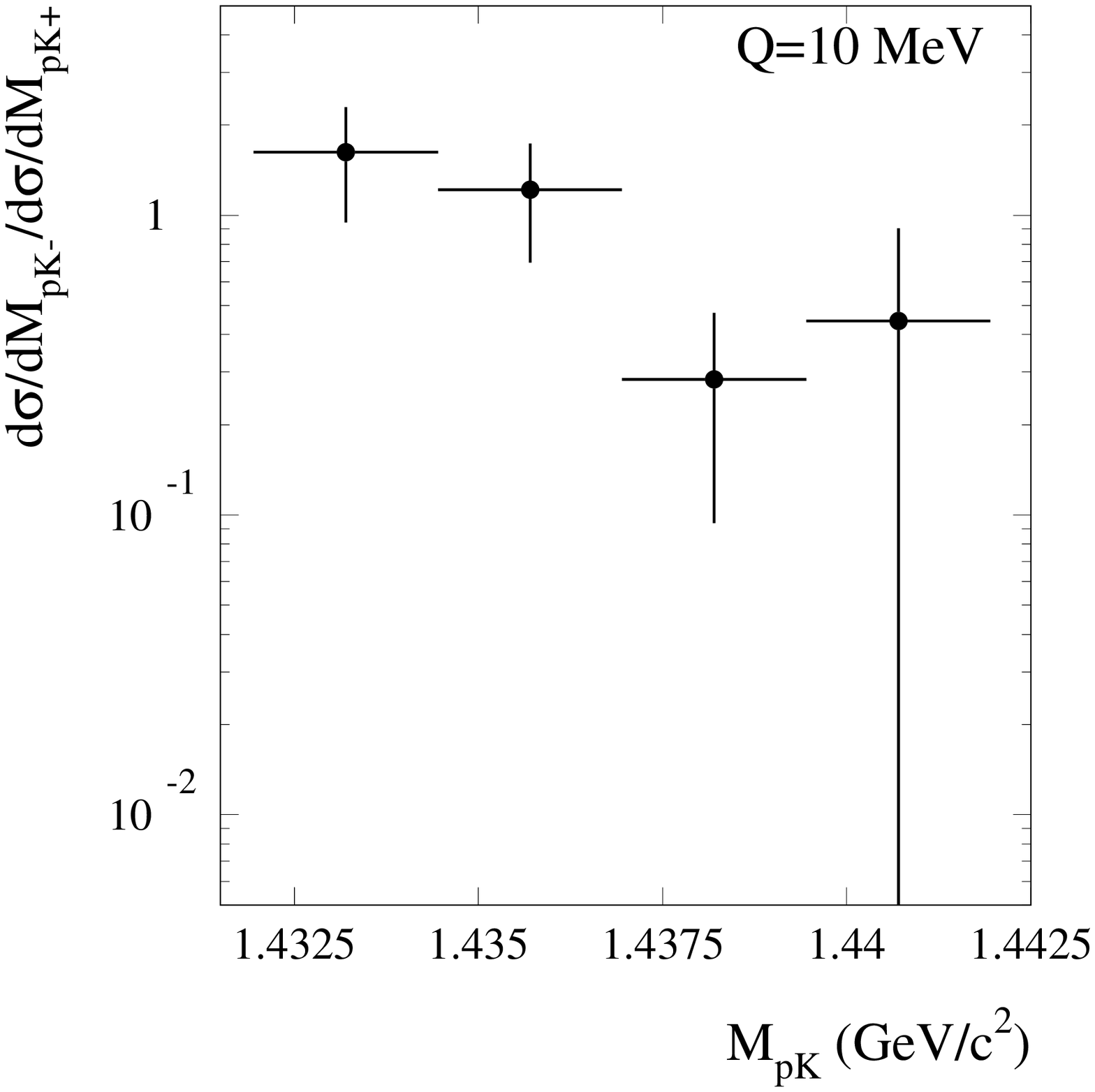}
\includegraphics[width=0.49\textwidth]{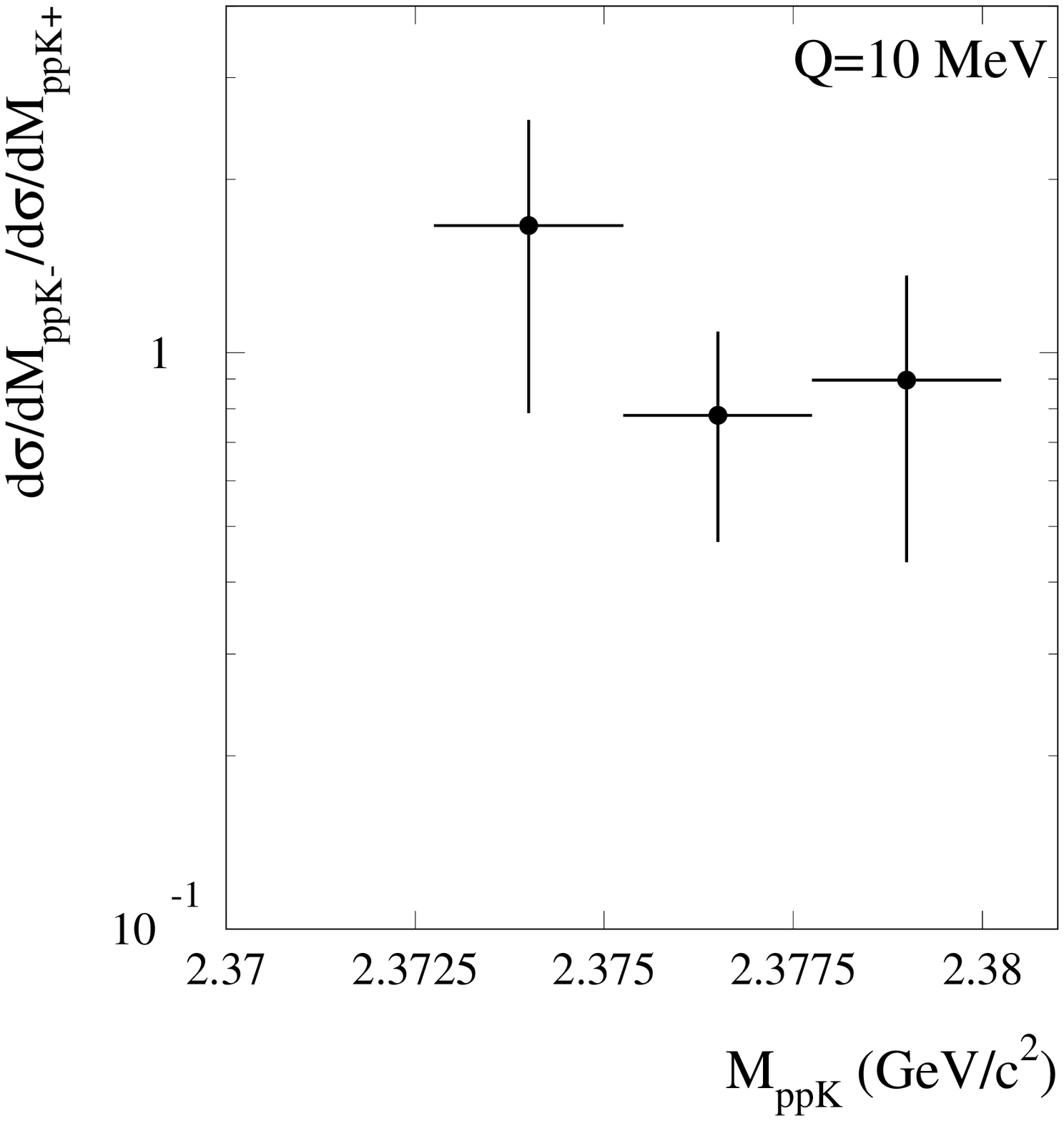}
\includegraphics[width=0.49\textwidth]{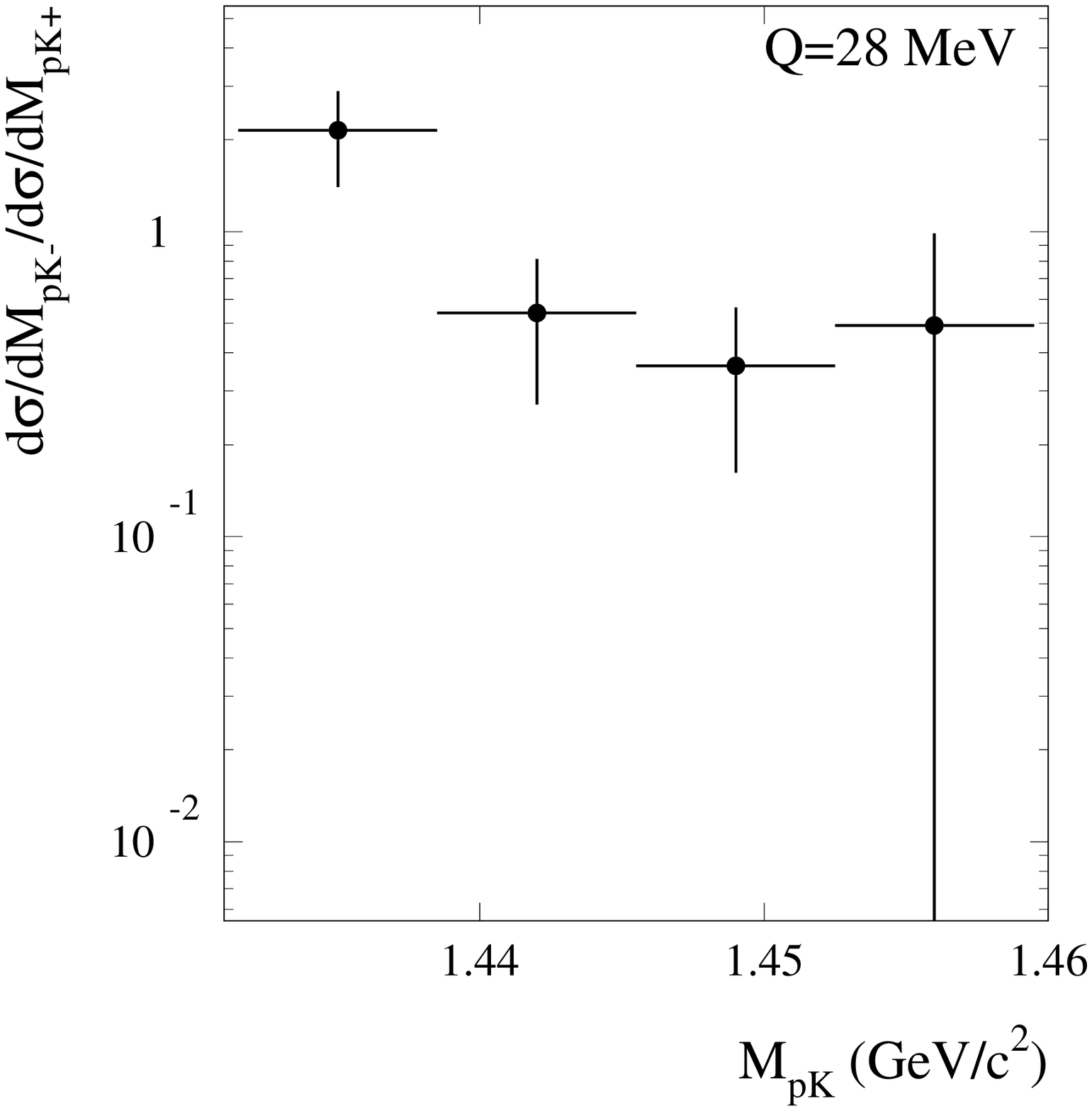}
\includegraphics[width=0.49\textwidth]{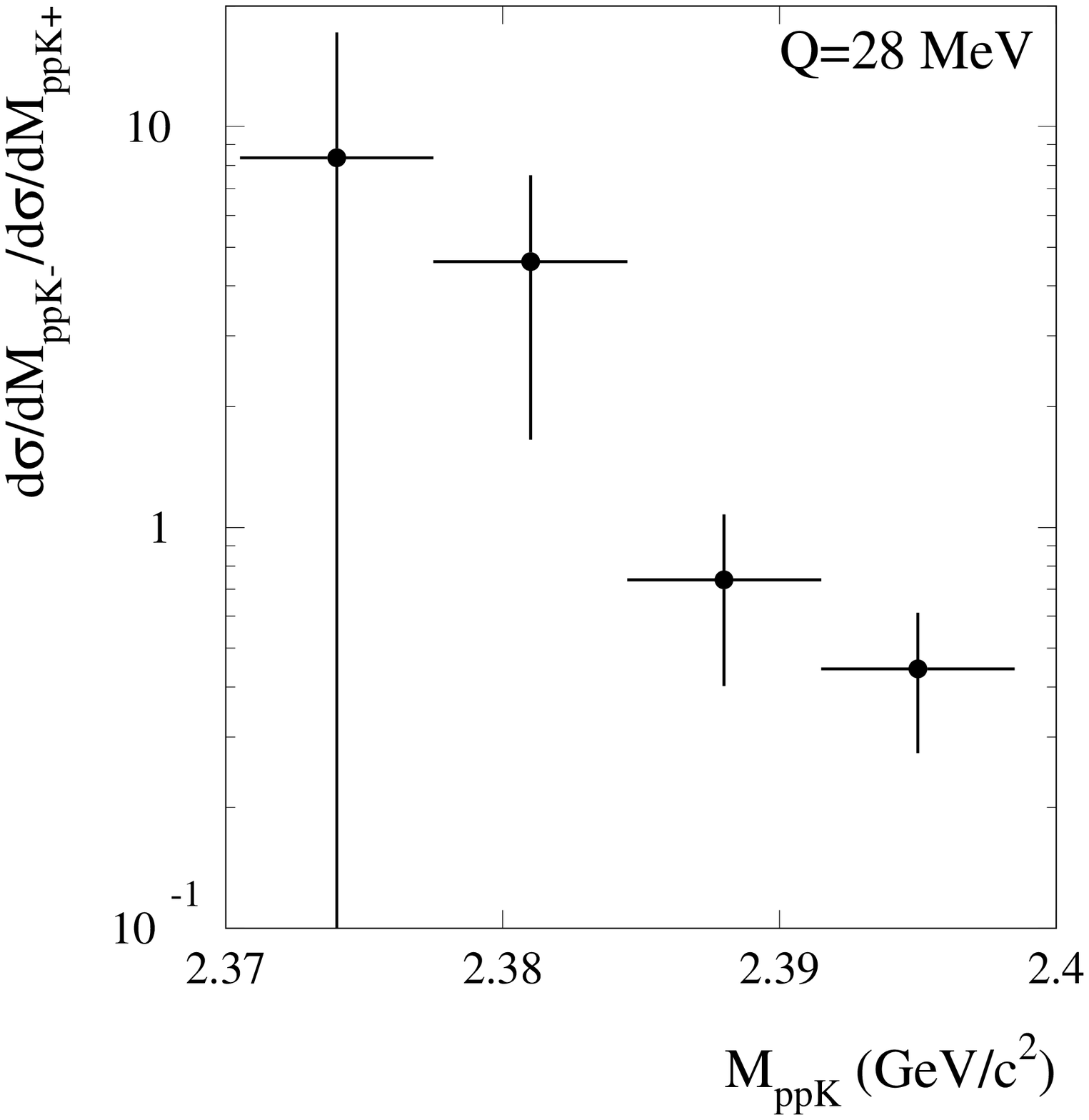}
\caption{Ratios of differential cross sections as a function of $pK^-$ invariant mass ($M_{pK^{-}}$)
and $ppK^{-}$ invariant mass ($M_{ppK^{-}}$) measured by the COSY-11 experiment at excess energies
of Q=10 MeV and Q=28 MeV~\cite{winter,PhysRevC}.}
\label{ratioC}
\end{figure}
These measurements revealed however that the total cross sections for this reaction near threshold are
in the order of nanobarns making these studies difficult due to low statistics~\cite{winter,wolke,quentmeier}.
Moreover, the possible $f_{0}$ or $a_{0}$ signal was too weak to be observed with COSY-11 in
the proton-proton collisions~\cite{quentmeier,jphysg}.
However, COSY-11 data showed unambiguous signs of the $pK^-$ final state interaction. It manifested itself
particularly strongly in the $pK^-$ and $ppK^-$ invariant mass distributions measured at excess energies of
Q=~10~MeV and Q=~28~MeV. The following ratios:
\begin{eqnarray}
\nonumber
R_{pK} &= \frac{\mathrm{d}\sigma/\mathrm{d}M_{pK^{-}}}{\mathrm{d}\sigma/\mathrm{d}M_{pK^{+}}}~,\\
\nonumber
R_{ppK} &= \frac{\mathrm{d}\sigma/\mathrm{d}M_{ppK^{-}}}{\mathrm{d}\sigma/\mathrm{d}M_{ppK^{+}}}~,
\end{eqnarray}
showed a significant enhancement in the region of both low $pK^-$ invariant mass $M_{pK^{-}}$,
and the low $ppK^-$ invariant mass $M_{ppK^{-}}$~\cite{winter,PhysRevC} (see Fig.~\ref{ratioC}).
Since the $pK^+$ interaction is known to be very weak this enhancement indicates a strong influence
of the $pK^-$ final state interaction. This effect has been observed then also by the ANKE
collaboration at higher energies with data of much better statistics~\cite{anke,Ye,anke_last}.
Examples of $R_{pK}$ and $R_{ppK}$ distributions measured by the ANKE collaboration
are presented in Fig.~\ref{ratioA}.
The influence of final state interaction in the low energy $ppK^+K^-$ system manifests itself
also in the shape of the $pp \to ppK^+K^-$ excitation function, where one observes a strong
deviation from the pure phase space expectations.
%
%
\section{Description of the dynamics in the low energy $ppK^+K^-$ system}
\label{sec:2}
Since shapes of the ratios presented in the previous section indicated a strong $pK^-$ attraction,
the $pp \to ppK^+K^-$ reaction near threshold was described in terms of the final state interaction.
Because we are dealing with the close-to-threshold region the complete transition matrix element for
this reaction may be factorized approximately as~\cite{moskal}:
\begin{equation}
\left|M_{pp\to ppK^+K^-}\right|^2 \approx \left| M_{0}\right|^2\cdot \left| M_{FSI}\right|^2~,
\end{equation}
where $\left| M_{0}\right|^2$ represents the total short range production amplitude, and $\left| M_{FSI}\right|^2$
denotes the final state interaction enhancement factor. The ANKE collaboration proposed a simple
ansatz assuming factorization of $M_{FSI}$ to the two-particle scattering amplitudes~\cite{wilkin},
taking into account strong proton-proton and $pK^-$ interactions and neglecting the $K^+$
influence\footnote{This is a very rough approximation, but more realistic calculations for four-body final
states are not available.} :
\begin{equation}
M_{FSI} = F_{pp}(k_{1}) \times F_{p_{1}K^-}(k_{2}) \times F_{p_{2}K^-}(k_{3})~,
\label{row1}
\end{equation}
where $k_{1}$, $k_{2}$ and $k_{3}$ denote the relative momentum of particles in the proton-proton and
two proton-$K^-$ subsystems.
Using this approximation one can describe well all the measured differential distributions using an
effective scattering length $a_{pK^-} = i$1.5~fm~\cite{anke}.\\
This model, however, underestimates COSY-11 total cross sections near threshold, which indicates
that in the low energy region the influence of the $K^{+}K^{-}$ final state interaction may be significant.
Motivated by this observation the COSY-11 collaboration has performed analysis of the low
energy $pp \to ppK^{+}K^{-}$ Goldhaber Plot distributions measured at excess energies of Q~=~10~MeV
and 28~MeV~\cite{PhysRevC}.
The final state interaction model used in that analysis was based on the factorization
ansatz in Eq.~\ref{row1}, with an additional term describing the interaction of the $K^+K^-$
pair.
The proton--proton scattering amplitude was taken into account using the following parametrization:
\begin{equation}
F_{pp} =
  \frac{e^{i\delta_{pp}({^{1}\mbox{\scriptsize S}_{0}})} \cdot
        \sin{\delta_{pp}({^{1}\mbox{S}_0})}}
       {Ck_{1}}~,
\label{ppfsi}       
\end{equation}
where $C$ stands for the square root of the Coulomb pe\-ne\-tra\-tion factor~\cite{pp-FSI}.
The parameter $\delta_{pp}({^{1}\mbox{S}_0})$ denotes the phase shift 
calculated according to the modified Cini--Fubini--Stanghellini formula with
the Wong--Noyes Coulomb correction~\cite{noyes995,noyes465,naisse506}.
Moreover, factors describing the enhancement originating from the $pK^-$ and $K^+K^-$--FSI were
parametrized using the scattering length approximation:
\begin{equation}
\nonumber
F_{pK^{-}}=\frac{1}{1-ika_{pK^-}}~,~~~F_{K^{+}K^{-}}=\frac{1}{1-ik_{4}~a_{K^+K^-}}~,
\label{F_ppKK}
\end{equation}
where $a_{pK^-} = i$1.5 fm  and $a_{K^+K^-}$ is the scattering length of the $K^{+}K^{-}$
interaction treated as a free parameter in the analysis. 
As a result of these studies $a_{K^+K^-}$ was estimated to be:
$\left|Re(a_{K^{+}K^{-}})\right| = 0.5^{~+4}_{~-0.5}$~fm and $Im(a_{K^{+}K^{-}}) = 3~\pm~3$~fm.\\
This model neglects any coupled channel effects, like e.g. the charge-exchange
interaction allowing for the $K^{0}\bar{K^{0}}\rightleftharpoons K^{+}K^{-}$ transitions or rescattering
to scalar mesons: $K^+K^- \to f_0(980)/a_0(980) \to K^+K^-$, which would
generate a significant cusp effect in the $K^+K^-$ invariant mass spectrum near the $K^0\bar{K^0}$
threshold~\cite{dzyuba}, and the $a_{K^+K^-}$ isospin dependence. The detailed analysis of the
$K^+K^-$ invariant mass distributions measured by the ANKE experiment showed however, that these
effects cannot be distinguished
from the pure kaons elastic scattering and the production with isospin I~=~0 is dominant in the
$pp\to ppK^+K^-$ reaction independently on the exact values of the scattering lengths~\cite{dzyuba}.\\
Since the shape of the excitation function for the $pp\to ppK^+K^-$ reaction appeared to be
quite sensitive to the final state interaction in the close-to-threshold region, we have
extended the analysis of differential cross sections measured by the COSY-11 collaboration
at Q~=~10 and Q~=~28 MeV taking into account in the fit also all the $pp\to ppK^+K^-$
total cross sections measured near threshold~\cite{eqcd13,physRevC2}. Moreover, since
the $pK^{-}$ scattering length estimated by the ANKE group is rather an effective
parameter~\cite{anke}, in this analysis we have used more realistic $a_{pK^-}$ value estimated
independently as a mean of all the scattering length values summarized in Ref.~\cite{Yan:2009mr}:
$a_{pK^-} = (-0.65 + 0.78i$)~fm. The energy range for the experimental excitation function
is rather big, thus the $K^+K^-$ final state enhancement factor was parametrized using the effective
range expansion:
\begin{equation}
\nonumber
F_{K^{+}K^{-}}=\frac{1}{\frac{1}{a_{K^+K^-}}+\frac{b_{K^+K^-}k^2_{4}}{2}-ik_{4}},
\label{F_KKb}
\end{equation}
where $a_{K^+K^-}$ and $b_{K^+K^-}$ are the scattering length and the effective range
of the $K^{+}K^{-}$ interaction, respectively.
As a result of these studies we have obtained the following values of $K^+K^-$
final state interaction parameters~\cite{physRevC2}:
\begin{eqnarray}
\nonumber
\mathrm{Re}(b_{K^{+}K^{-}}) = -0.1 \pm 0.4_{stat} \pm~0.3_{sys}~\mathrm{fm}\\
\nonumber
\mathrm{Im}(b_{K^{+}K^{-}}) = 1.2^{~+0.1_{stat}~+0.2_{sys}}_{~-0.2_{stat}~-0.0_{sys}}~\mathrm{fm}\\
\nonumber
\left|\mathrm{Re}(a_{K^{+}K^{-}})\right| = 8.0^{~+6.0_{stat}}_{~-4.0_{stat}}~\mathrm{fm}\\
\nonumber
\mathrm{Im}(a_{K^{+}K^{-}}) = 0.0^{~+20.0_{stat}}_{~-5.0_{stat}}~\mathrm{fm}~.
\nonumber
\label{chi2resultsB}
\end{eqnarray}
The fit is in principle sensitive to both the scattering length and effective range,
however, with the available low statistics data the sensitivity to $a_{K^{+}K^{-}}$ is very weak.
\section{Update of the COSY-11 total cross sections measured at Q~=~6~MeV and Q~=~17~MeV}
\begin{table}
\begin{center}
\begin{tabular}{|c|c|c|}
\hline
 Q [MeV] & $\sigma_{old} \mathrm{[nb]} $ & $\sigma_{new} \mathrm{[nb]} $\\
\hline
6 & 0.49 $\pm$ 0.40  & 0.51 $\pm$ 0.42\\
\hline
17 & 1.80 $\pm$ 0.27 & 1.88 $\pm$ 0.28\\
\hline
\end{tabular}
\end{center}
\caption{
\label{tab1}
Total cross sections measured by the COSY-11 experiment at excess energies of
Q~=~6~MeV and Q~=~17~MeV determined with the old values of luminosity ($\sigma_{old}$) and taking into
account the newest EDDA cross sections ($\sigma_{new}$).}
\end{table}
In all the COSY-11 measurements of the $pp\to ppK^+K^-$ reaction the luminosity needed for evaluation
of cross sections was determined based on the simultaneous registration of elastically scattered protons.
The differential counting rates of elastic protons scattering measured together with the $pp\to ppK^+K^-$
reaction were then compared to data obtained by the EDDA collaboration.
The luminosity for measurements at Q~=~6~MeV and Q~=~17~MeV was calculated using EDDA data
gathered in 1997~\cite{edda1}, while for measurements at the two other excess energies the updated
and much more precise EDDA differential cross sections were used~\cite{edda2}. Therefore, we have
reevaluated the COSY-11 luminosities at Q~=~6~MeV and Q~=~17~MeV which resulted in new total cross
section values for these excess energies~\cite{eryk_acta}. The updated values of the cross sections are gathered in
Tab.~\ref{tab1}. One can see that they are slightly higher than the old published total cross
sections~\cite{wolke,quentmeier} which increases the observed enhancement at threshold.
Therefore, it is worth to check how the values of scattering length and effective range
obtained in~\cite{physRevC2} change for a fit which takes into account the updated COSY-11
cross sections.
\section{Determination of the $K^+K^-$-FSI parameters taking into account updated COSY-11 cross sections}
\begin{figure}
\centering
  \includegraphics[width=0.49\textwidth,angle=0]{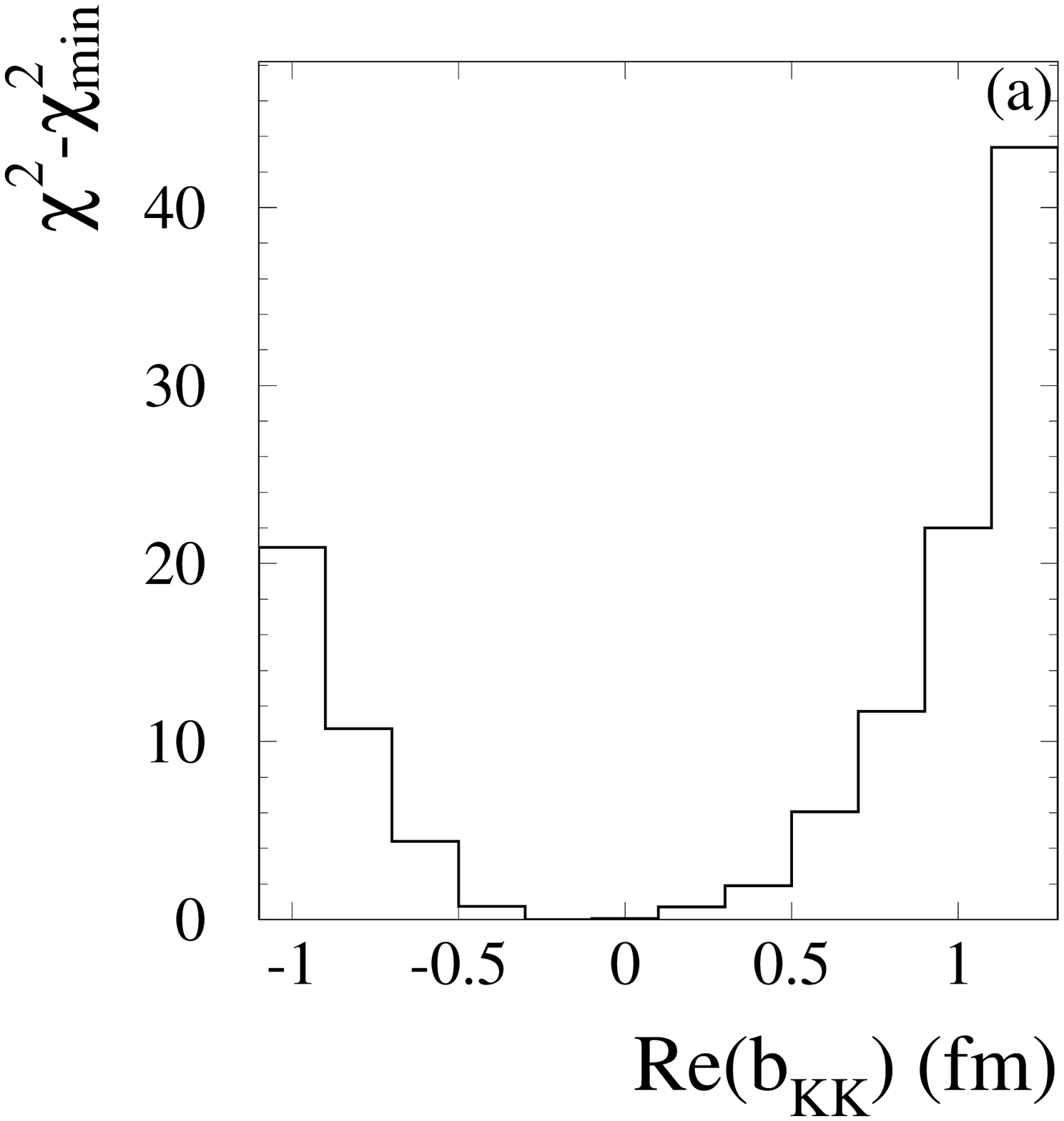}
 \includegraphics[width=0.49\textwidth,angle=0]{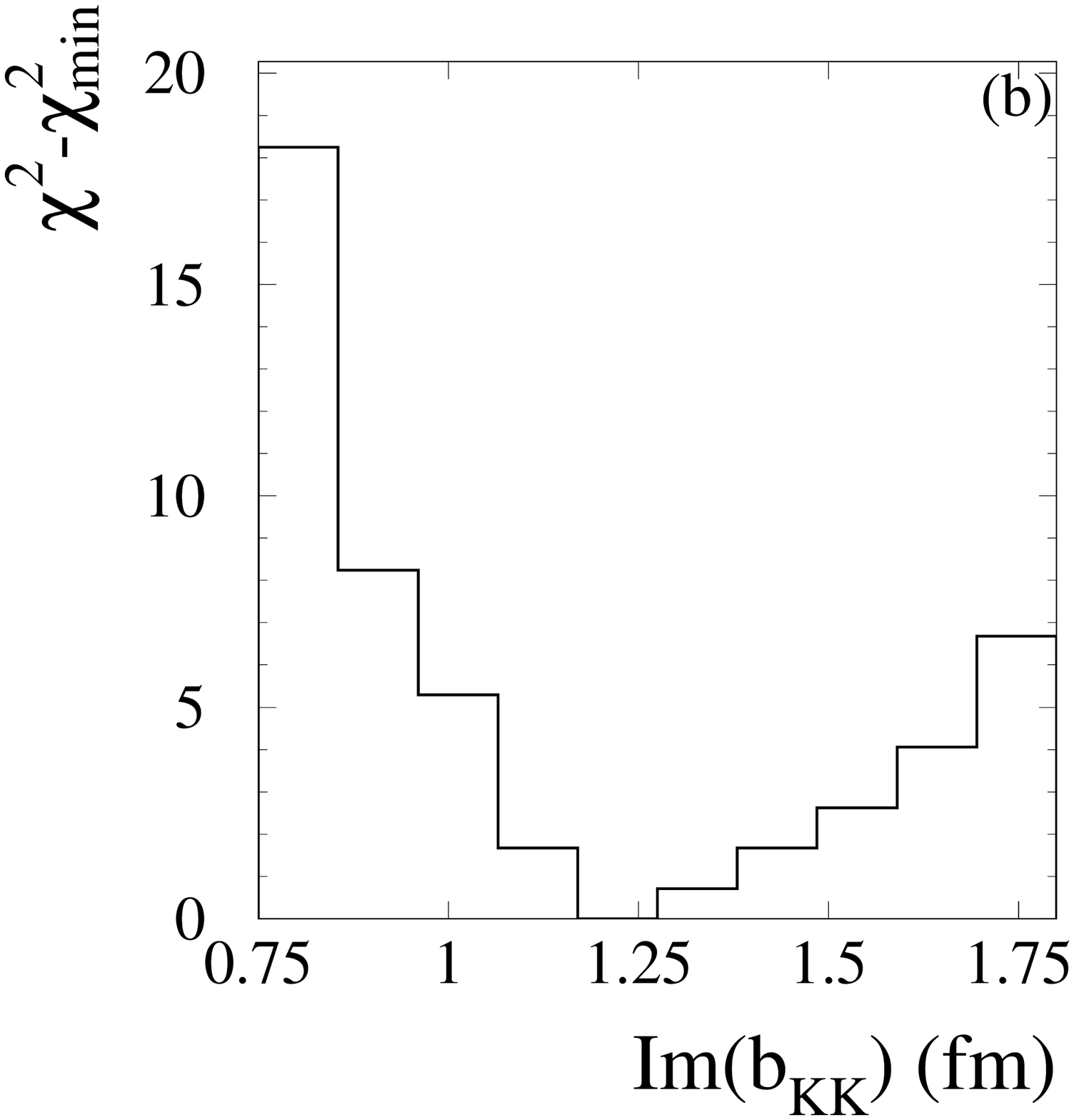}
  \includegraphics[width=0.49\textwidth,angle=0]{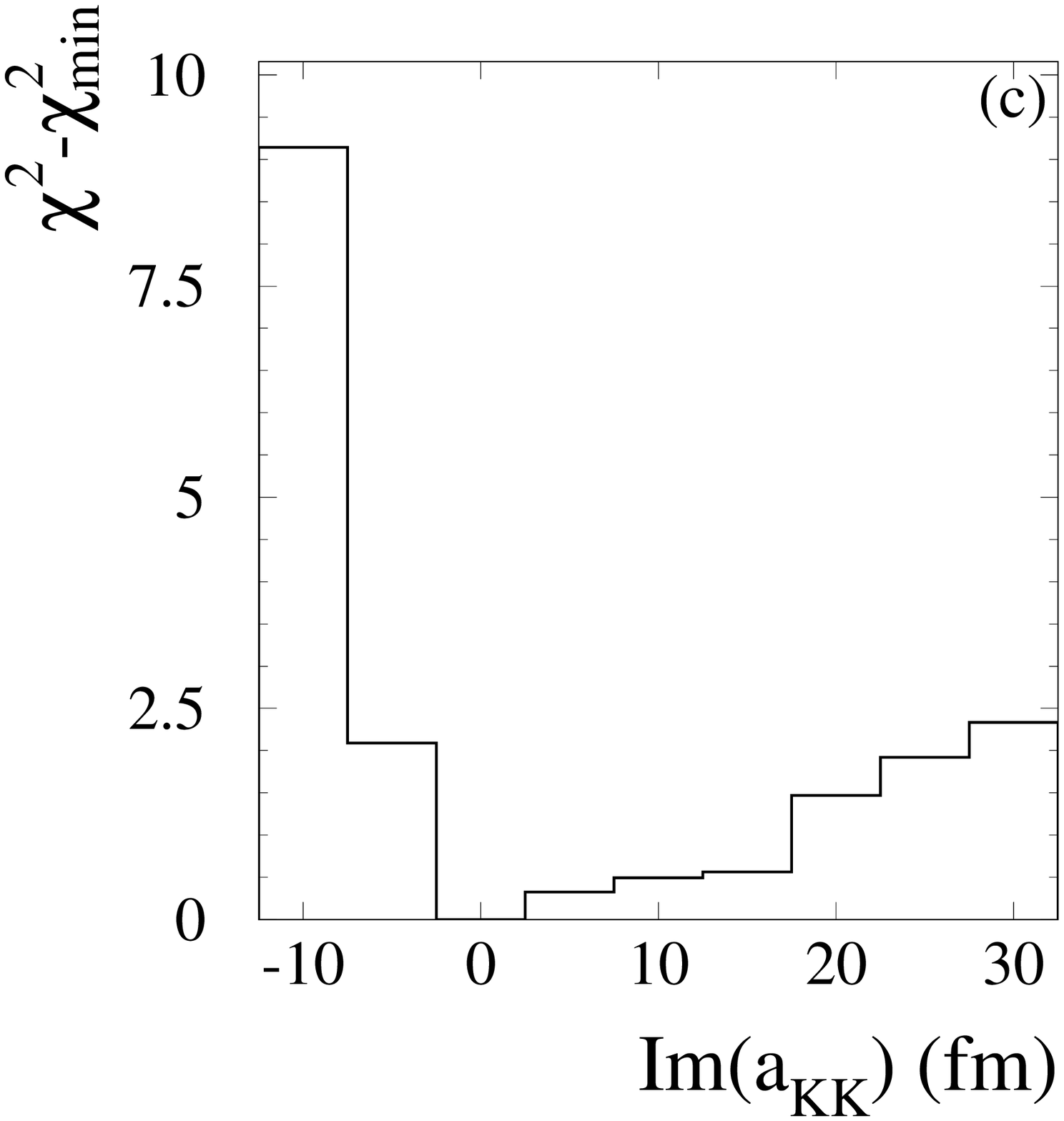}
  \includegraphics[width=0.49\textwidth,angle=0]{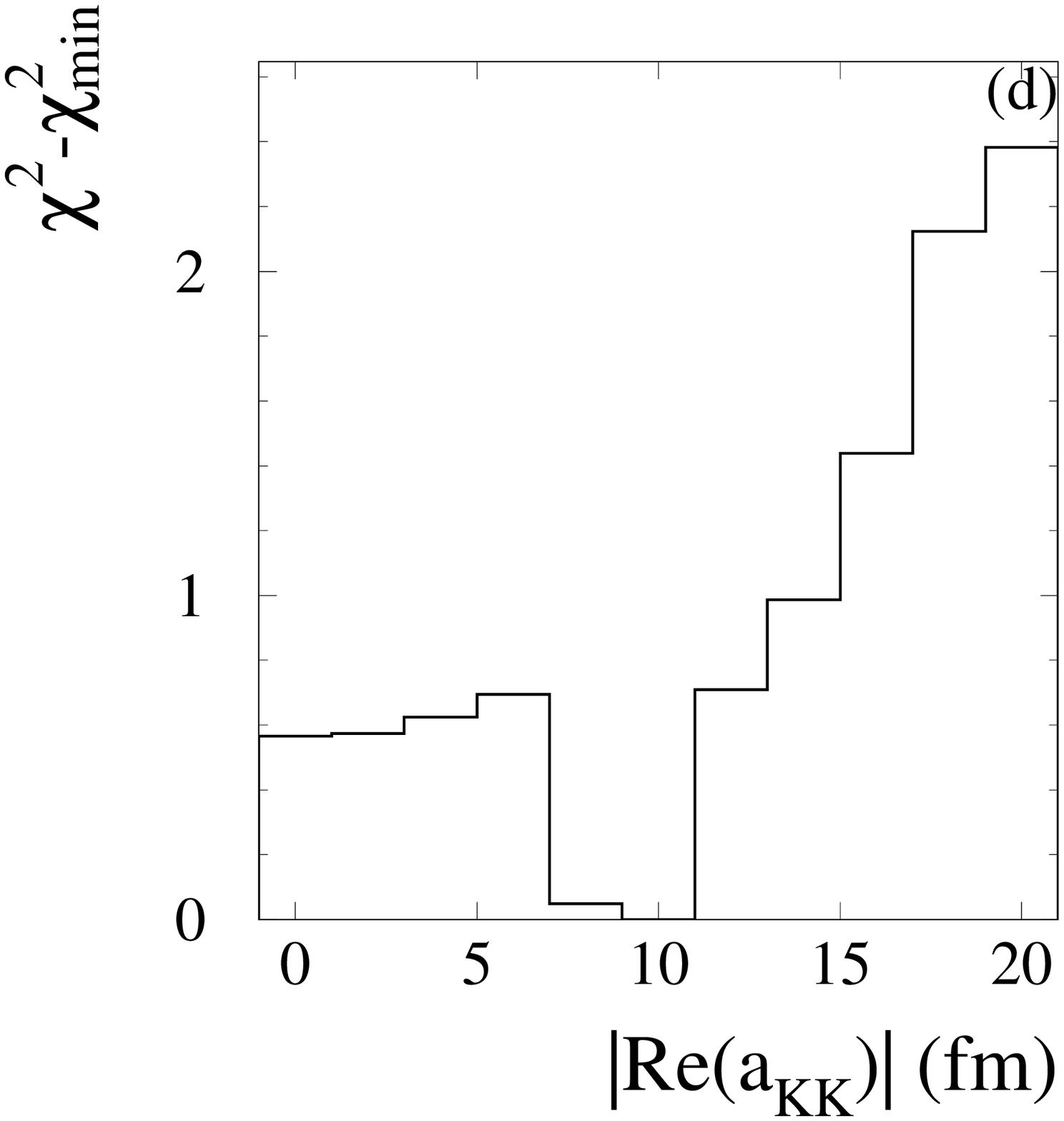}
\caption{
$\chi^2-\chi^2_{min}$ distribution as a function of:
(a) $\mathrm{Re}(b_{K^+K^-})$, (b) $\mathrm{Im}(b_{K^+K^-})$,
(c) $\mathrm{Im}(a_{K^{+}K^{-}})$ and (d) $|\mathrm{Re}(a_{K^{+}K^{-}})|$.
$\chi^{2}_{min}$ denotes the absolute minimum with respect to parameters
$\alpha$, $\mathrm{Re}(b_{K^+K^-})$, $Im(b_{K^+K^-})$, $|\mathrm{Re}(a_{K^+K^-})|$,
and $Im(a_{K^+K^-})$.
}
\label{fig:2}
\end{figure}
In the new fit we have taken into account not only the updated \mbox{COSY-11} cross section but also the newest
measurement of the ANKE group done at Q~=~24~MeV~\cite{anke_last}. As in the previous analysis~\cite{physRevC2}
we have preformed combined fit to Goldhaber plots measured at excess energies of Q~=~10~MeV and Q~=28~MeV
and to the excitation function determined near the threshold.
To determine $a_{K^{+}K^{-}}$ and
$b_{K^+K^-}$ we have constructed the following $\chi^{2}$ statistics:
\begin{eqnarray}
\nonumber
\chi^2\left(a_{K^+K^-},b_{K^+K^-},\alpha\right) = \sum_{i=1}^{8}\frac{\left(\sigma_{i}^{expt}
- \alpha\sigma_{i}^{m}\right)^2} {\left(\Delta\sigma_{i}^{expt}\right)^2}\\
 +2\sum_{j=1}^{2}\sum_{k=1}^{10} \, [\beta_{j} N_{jk}^s - N_{jk}^e +  N_{jk}^e \,
{\mathrm{ln}}(\frac{N_{jk}^e}{\beta_{j} N_{jk}^s})] ,
\label{eqchi2_mh}
\end{eqnarray}
where the first term was defined following the Neyman's $\chi^{2}$ statistics, and accounts for the excitation
function near the threshold for the $pp \to ppK^{+}K^{-}$ reaction.
$\sigma_{i}^{expt}$ denotes the $i$th experimental total cross section
measured with uncertainty $\Delta\sigma_{i}^{expt}$ and $\sigma_{i}^{m}$
stands for the calculated total cross section normalized with a factor $\alpha$
which is treated as an additional parameter of the fit.
$\sigma_{i}^{m}$ was calculated for each excess energy Q as a phase space integral
over five independent invariant masses~\cite{nyborg}:
\begin{eqnarray}
\nonumber
\sigma^{m}=\int\frac{\pi^{2}\left|M\right|^{2}}{8s\sqrt{-B}}~\mathrm{d}M^{2}_{pp}\mathrm{d}M^{2}_{K^{+}K^{-}}
\mathrm{d}M^{2}_{pK^{-}}\mathrm{d}M^{2}_{ppK^{-}}\mathrm{d}M^{2}_{ppK^{+}}.
\label{goldhaber3}
\end{eqnarray}
Here $s$ denotes the square of the total energy of the system determining the value
of the excess energy, and $B$ is a function of the invariant masses with the exact
form to be found in Nyborg's work~\cite{nyborg}.\\
The amplitude for the process $|M|^2$ contains the FSI enhancement factor defined in Eq.~(\ref{row1}) 
with additional factor expressing the $K^+K^-$ interaction. The $pp$, $pK^-$ and $K^+K^-$
interactions were parametrized according to Eq.~\ref{ppfsi}, Eq.~\ref{F_ppKK} and Eq.~\ref{F_KKb},
respectively.
The second term of Eq.~(\ref{eqchi2_mh}) corresponds to the Poisson likelihood
chi--square value~\cite{baker} describing the fit to the Goldhaber plots.
$N_{jk}^e$ denotes the number of events in the $k$th bin of the $j$th experimental
Goldhaber plot, and $N_{jk}^s$ stands for the content of the same bin in the simulated
distributions. $\beta_{j}$ is a normalization factor which is fixed by values of the fit parameters,
and which is defined for the $j^{th}$ excess energy as the ratio of the total number of events expected
from the calculated total cross section $\sigma_{j}^{m}$ and the total luminosity
$L_{j}$~\cite{winter}, to the total number of simulated $pp \to ppK^+K^-$ events $N_{j}^{gen}$:
\begin{eqnarray}
\nonumber
\beta_{j} = \frac{L_{j}\alpha\sigma_{j}^{m}}{N_{j}^{gen}}.
\label{beta}
\end{eqnarray} 
The $\chi^2$ distributions (after subtraction of the minimum value) are presented
as a function of the real and imaginary parts of $a_{K^+K^-}$ and $b_{K^+K^-}$ in Fig.~\ref{fig:2}. 
The best fit to the experimental data corresponds to:
\begin{eqnarray}
\nonumber
\mathrm{Re}(b_{K^{+}K^{-}}) = -0.2^{+0.8_{stat}~+0.4_{sys}}_{-0.6_{stat}~-0.4_{sys}}~\mathrm{fm}\\
\nonumber
\mathrm{Im}(b_{K^{+}K^{-}}) = 1.2^{~+0.5_{stat}~+0.3_{sys}}_{~-0.3_{stat}~-0.3_{sys}}~\mathrm{fm}\\
\nonumber
\left|\mathrm{Re}(a_{K^{+}K^{-}})\right| = 10^{~+17_{stat}}_{~-10_{stat}}~\mathrm{fm}\\
\nonumber
\mathrm{Im}(a_{K^{+}K^{-}}) = 0^{~+37_{stat}}_{~-10_{stat}}~\mathrm{fm}~,\\
\nonumber
\label{chi2resB}
\end{eqnarray}
with a $\chi^2$ per degree of freedom of: $\chi^2/ndof = 1.70$. The statistical uncertainties
in this case were determined at the 70\% confidence level taking into account that we have varied
five parameters~\cite{james}.
\begin{figure}
\centering
\includegraphics[width=0.6\textwidth]{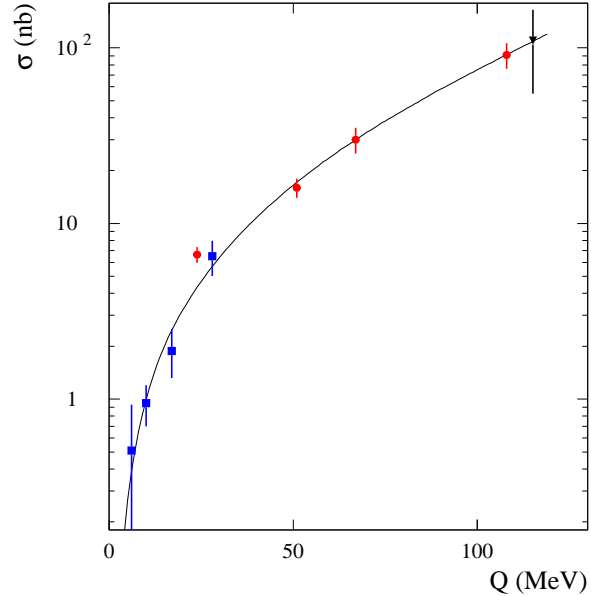}
\caption{Excitation function for the $pp\rightarrow ppK^{+}K^{-}$ reaction.
Triangle and circles represent the DISTO and ANKE measurements, respectively~\cite{anke,anke_last,disto}.
The squares are results of the COSY--11~\cite{wolke,quentmeier,PhysRevC} measurements.
The solid curve corresponds to the result of calculations obtained taking into account
$pp$, $pK^-$, and $K^+K^-$ interactions using the scattering length $a_{K^{+}K^{-}}$ and effective
range $b_{K^+K^-}$ obtained in the fit taking into account updated COSY-11 cross sections and the latest
ANKE measurement.}
\label{funkcja_wzbudzenia}
\end{figure}
As in the previous analysis we have estimated also systematic errors due to the assumed $pK^{-}$
scattering length by repeating the analysis for every $a_{pK^-}$ value quoted  in Ref.~\cite{Yan:2009mr}.
Due to the fact that in the case of scattering length the obtained systematic uncertainties are much
smaller than the statistical ones we neglect them in the final result.
\section{Summary and outlook}
The new analysis of the $K^+K^-$ final state interaction performed with updated
COSY-11 cross sections and taking into account the latest ANKE measurement resulted
in the new estimates of the $K^+K^-$ scattering length and effective range. As in the
previous analysis the fit is in principle sensitive to the effective range and with
the available low statistics the sensitivity to the scattering length is very weak.\\
The latest ANKE results obtained at Q~=~24 MeV suggest however, that for the more
accurate description of the interaction in the $ppK^+K^-$ system a much more
sophisticated model than the factorization ansatz used so far is needed~\cite{anke_last}.
Thus, the results of analysis quoted in this article should be considered
rather as effective parameters.
\section{Acknowledgements}
The author is grateful to P.~Moskal and E.~Czerwi{\'n}ski for their valuable comments
and corrections and for providing updated values of the \mbox{COSY-11} luminosities
for measurements at Q~=~6 MeV and Q~=~17 MeV. This research was supported by the FFE
grants of the Research Center J{\"u}lich, by the Polish National Science Center and
by the Foundation for Polish Science.

\end{document}